\documentclass[fleqn,usenatbib]{mnras}

\usepackage{newtxtext,newtxmath}

\usepackage[T1]{fontenc}
\usepackage{ae,aecompl}

\usepackage{graphicx}	
\usepackage{amsmath}	
\usepackage{amssymb}	
\usepackage{pdflscape}
\usepackage{rotating}

\newcommand{\eg}{e.g.,\ }

\newcommand{\Msun}{$M_{\odot}$}

\newcommand{\kms}{km~s$^{-1}$}

\newcommand{\OI}{O~{\sc i}}

\newcommand{\CaII}{Ca~{\sc ii}}

\newcommand{\FeII}{Fe~{\sc ii}}

\newcommand{\CoII}{Co~{\sc ii}}

\newcommand{\NiII}{Ni~{\sc ii}}

\newcommand{\Nifs}{$^{56}$Ni}
\newcommand{\Mej}{$M_{\rm ej}$}
\newcommand{\KE}{$E_{\rm kin}$}

\newcommand{\ab}{$\sim$}

\title{Extracting high-level information from gamma-ray burst supernova spectra}

\author[C. Ashall et al.]{
\parbox{\textwidth}{
\raggedright
C.Ashall$^{1}$, $\thanks{E-mail:chris.ashall24@gmail.com}$
P.A. Mazzali$^{2,3}$, 
}\vspace{0.4cm}\\
\parbox{\textwidth}{
$^{1}$Department of Physics, Florida State University, Tallahassee, FL 32306, USA\\
$^{2}$Astrophysics Research Institute, Liverpool John Moores University, IC2, Liverpool Science Park, 146 Brownlow Hill, Liverpool L3 5RF, UK\\
$^{3}$Max-Planck-Institut f\"ur Astrophysik, Karl-Schwarzschild-Str. 1, D-85748 Garching, Germany\\
}
\vspace{-0.75cm}
}
\date{Accepted 2020 January 21. Received 2020 January 21; in original form 2019 April 02}

\pubyear{2017}

\begin{document}
\label{firstpage}

\pagerange{\pageref{firstpage}--\pageref{lastpage}}
\maketitle

\begin{abstract}
Radiation transport codes are often used in astrophysics 
to construct spectral models. 
In this work we demonstrate how producing these models 
for a time series of data can provide unique information about supernovae (SNe). 
Unlike previous work, we specifically concentrate on the method for obtaining the best synthetic spectral fits, 
and the errors associated with the preferred model parameters. 
We demonstrate how varying the ejecta mass,
bolometric luminosity ($L_{bol}$) and photospheric  velocity ($v_{ph}$), 
affects the outcome of the synthetic spectra. 
As an example we analyze the photospheric phase spectra of the GRB-SN\,2016jca. 
It is found that for most epochs (where the afterglow subtraction is small)
the error on $L_{bol}$ and $v_{ph}$ was \ab5\%. 
The uncertainty on ejecta mass and \KE\ was found to be \ab20\%, although this can be expected 
to dramatically decrease if models of nebular phase data can be simultaneously produced. 
We also demonstrate how varying the elemental abundance in the ejecta can produce better 
synthetic spectral fits. In the case of SN\,2016jca it is found that a 
decreasing \Nifs\ abundance as a function of decreasing velocity 
produces the best fit models. This could be the case if the \Nifs\ was sythesised  at the side of
the GRB jet, or dredged up from the centre of the explosion. 
The work presented here can be used as a guideline for future studies on  supernovae
which use the same or similar radiation transfer code.
\end{abstract}

\begin{keywords}
Supernova, radiative transfer
\end{keywords}



\section{Introduction}

Radiation transfer codes are frequently used in astrophysics to obtain meaningful insight from observations. 
These codes usually fall in two categorizes: i) those which are used to fit individual 
observations, these codes are usually not computationally expensive, and ii) and those which are predictive but too computationally expensive to be run for every set of observations. 

In time domain astrophysics both types of codes have been throughly used to model 
many different transient sources including: core collapse supernovae 
\citep[\eg][]{Hoflich91,Baron95,Fisher00,Mazzali00,Thomas11,Jerkstrand12}, thermonuclear supernovae 
\citep[\eg][]{Kasen06,Kromer09,Kerzendorf14,Ashall16,Hoeflich17,Goldstein18,Ashall18a}, and more recently 
kilonova \citep[\eg][]{Tanaka13,Smartt17,Bulla19}. However, previous work has 
concentrated on the results obtained from these models, and often neglected 
to explain the  exact modeling method, fitting procedure or associated errors.

One subset of core collapse SN which have often been spectroscopically moddeled are 
stripped envelope supernovae (SE-SNe). 
These are end result of massive stars that explode after loosing their H/He shells through wind or binary interaction \citep[\eg][]{Puls08,Eldridge13}. They spectroscopically come in three classes, 
type IIb SNe (SNe IIb) which 
are H and He rich, type IIb SNe (SNe Ib) which are H poor but He rich, and type Ic SNe (SNe Ic)
which are deprived of both H and He \citep{Filippenko97,Matheson01}.

A subset of SNe Ic, know as broad line SNe Ic (SN Ic-BL) have very broad features and high specific kinetic energies (\KE), and the most energetic of these events have been known to be associated with X-ray flashes \citep{Mazzali06,Mazzali08a,Tanaka09} and long gamma-ray bursts (LGRBs) \citep[\eg][]{Iwamoto98,Galama98,Nakamura01,Stanek03,Mazzali03,Ferrero06,Woosley06,Bufano12,Jin13,Schulze14,Modjaz16,Toy16,Ashall19}. 

To date there have only been 6 GRB-SNe, all z$<0.2$, with a high quality time series of data which is good enough for spectroscopic modelling (see \eg \citet{Iwamoto98,Mazzali03,Deng05,Mazzali06}). 
However, these papers usually just concentrate on the results and do not have a full detailed explanation of the method, analysis and errors. 
Therefore, in this paper we concentrate on  this, as an example we use the spectra of
SN\,2016jca presented in \citet{Ashall19}. 
We specifically concentrate on the spectra of the SN itself. We start by explaining our modeling technique (section 2) and method (section 3), then in section 4 we discuss our starting parameters, followed by a full error analysis in section 5, in section 6 the \Nifs\ distribution in the ejecta is discussed and finally in section 7 we summarize the results and how this relates to all similar models.

\begin{figure*}
\centering
\includegraphics[scale=0.3]{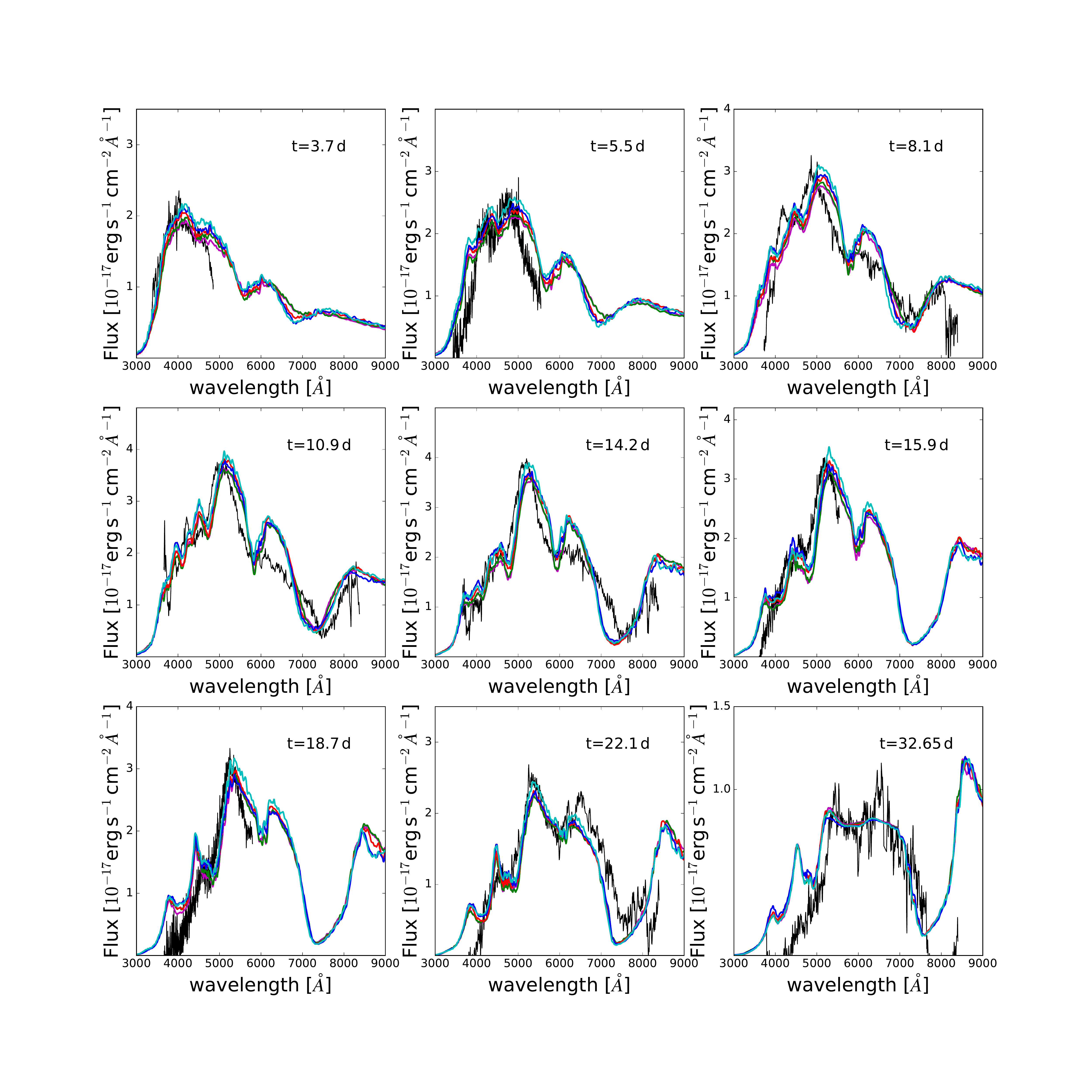}
\caption{Spectral models produced with five density profiles with the same shape but different masses. 
The masses of the models are 2\Msun (magneta), 4\Msun (green), 6.5\Msun (blue), 8\Msun (yellow), and 11\Msun (cyan) 
The black spectra are the observations.
The different models were produced to conserve $T(r)$.}
\label{fig:5masses}
\end{figure*}

\section{The code}
In this work we model 9 spectra of SN\,2016jca which were analysed in \citet{Ashall19}, they cover the time range (in rest-frame) from 3.7 to 32.65\,days past explosion. The spectra have been corrected for a GRB afterglow with a time of break of 13 days past explosion. 
Table \ref{table:logofmodel} contains
the individual epochs (in rest-frame) for each spectra, as well as the moddeling parameters which will be explained below.

\begin{figure*}
\centering
\includegraphics[scale=0.3]{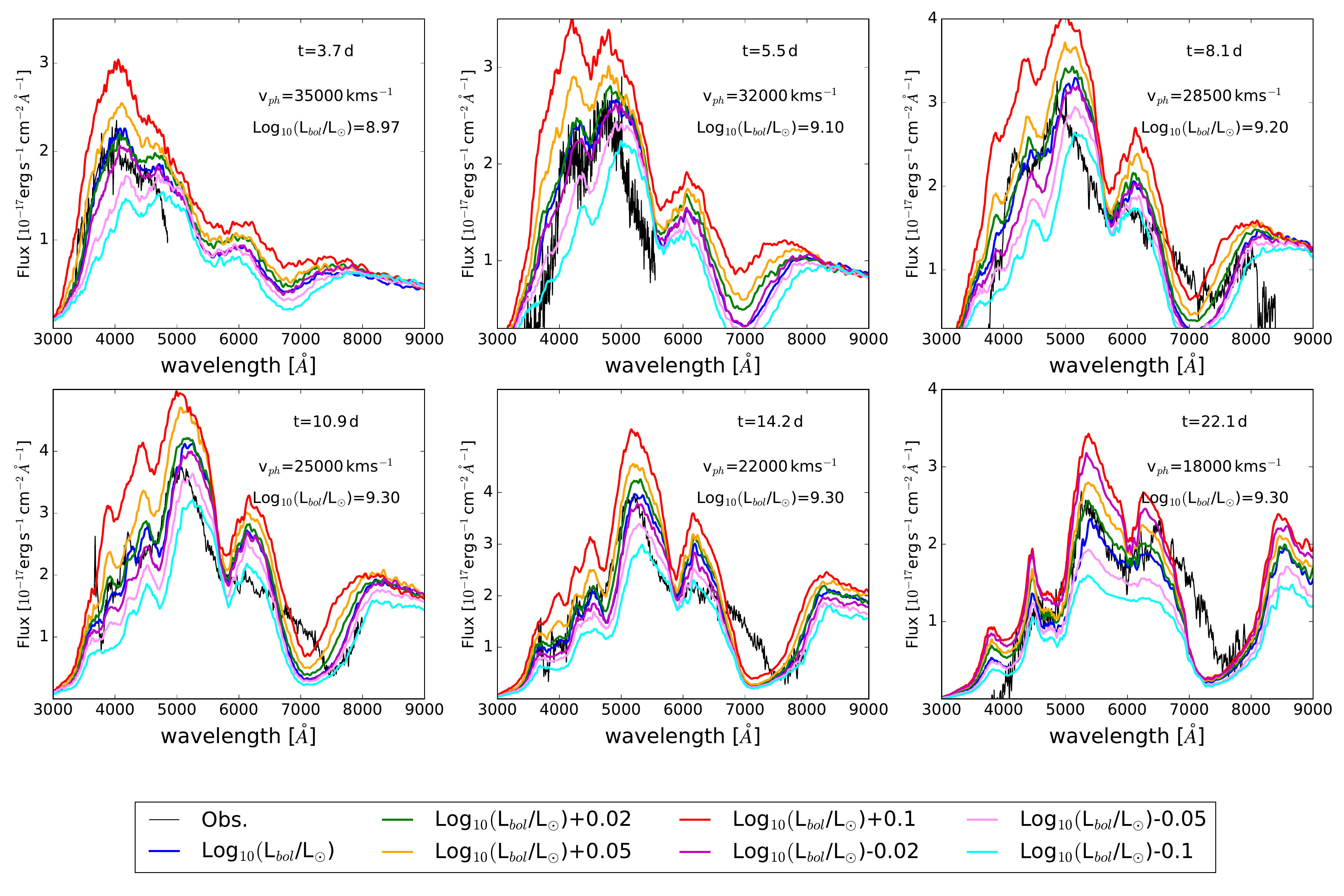}
\caption{Spectral observations and models of SN 2016jca with varying input 
luminosities. The blue model is the best fit model.}
\label{fig:varyL}
\end{figure*}

First we will discuss the code used in this analysis. 
A SN ejecta can be assumed to be in homologous starting a few minutes after explosion. 
This can be approximated by $r=v_{\rm ph} t_{\rm exp}$, where $r$ is the distance of the photosphere from the centre of the explosion,  $v_{\rm ph}$ is the photospheric 
velocity, and $t_{\rm exp}$ is the time from explosion. 
Therefore, as time passes deeper and deeper layers of the explosion can be observed.  

Here we use a 1D Monte Carlo radiative transfer code to produce synthetic spectra. 
The code, which is based on \citet{Mazzali93,Lucy99,Mazzali00}, follows the propagation of photon packets through a SN atmosphere. It has been used for many types of SNe before including SNe Ia \citep[\eg][]{Ashall14,Ashall16,Ashall18a,Galbany19} and SE-SN \citep[\eg][]{Prentice17,Mazzali17}. The code utilizes the 
Schuster-Schwarzschild approximation which assumes that 
the radiative energy is emitted at an inner boundary in the form of a black body. 
For GRB-SN this yields good results due to the amount of material above the photosphere.
Furthermore, the approximation does not require in-depth knowledge about the radiation  transport below the photosphere.

The photon packets have two fates: they either escape the ejecta or re-enter the photosphere, 
in a process known as back scattering. 
The photon packets can undergo Thomson scattering and line absorption in the ejecta. If a packet is absorbed,
the downward transition it follows is determined by a photon branching scheme, which allows florescence (blue to red) and 
reverse florescence (red to blue) to take place. The code utilises a modified nebular approximation to treat 
the excitation/ionization  state of the gas, to account for the non-local thermodynamical equilibrium effects caused by the diluted 
radiation field. There is an iteration between the radiation field and the state of the gas until convergence is achieved.
Finally, the formal integral of the radiation field is computed to obtain the spectrum.

When modelling SNe using this technique, the abundances of some elements (O, C, Ca, Mg, S, Si, Ti+Cr and \Nifs) are directly constrained in the data. Other elements (Neon being the most abundant element in this group) are present in explosion models but have no lines in the visible range covered by our data, so their abundances are fiducially assumed to match those of models. 
\\
The purpose of the code is to produce optimally fitting synthetic spectra. This is done by varying input parameters,
such as the bolometric luminosity ($L_{bol}$), photospheric velocity ($v_{\rm ph}$), and abundances ($Ab$). The code requires an input density profile, which can be scaled 
at the appropriate time from explosion for each modelled spectra, owing to the homologous expansion of the SN. We discuss the density profiles we have tested below.
\\
We have performed the analysis in 1D as carrying out the analysis in 2D or 3D
would have to rely entirely upon explosion
models, adding more free parameters, making computations much more expensive, and
most likely yield results that would only approximately match the observation, 
meaning we would most likely not get a more 
accurate solution.   We do note that if we were viewing 
the event perpendicular to the jet axis we would expect to 
see a different temperature and elemental abundance distribution than 
if we were viewing it down the axis of the GRB \citep{Tanaka07}.
Although our results are approximate, they should nevertheless be realistic if one
understands the constraints and limitations, as is shown by the fact that we can
fit the observations with physically sensible parameters. For example the abundance distributions derived using our radiation transport code for type Ia supernovae \citep{Ashall18a} yields very similar densities and abundance distributions to  those produced from non-local thermodynamic equilibrium radiation hydrodynamical models \citep{Hoeflich17}.

\begin{figure*}
\centering
\includegraphics[scale=0.3]{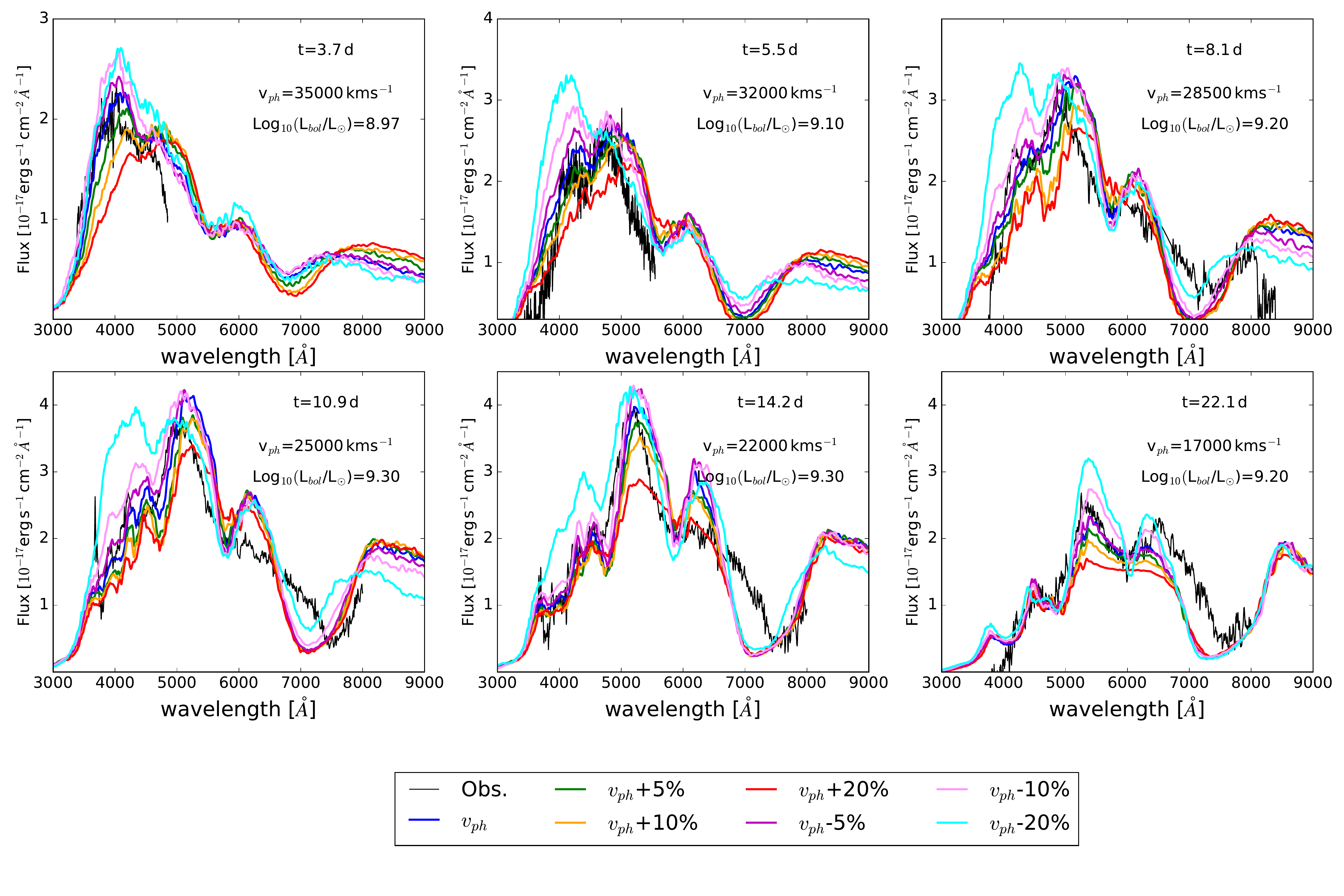}
\caption{Spectral observations and models of SN 2016jca with varying input 
photospheric velocities. The blue model is the best fit model.}
\label{fig:difV}
\end{figure*}

\section{Modelling method}
Our code requires as input a density distribution $\rho(r)$ (radius and velocity
are equivalent in a SN's homologously expanding ejecta, only a reference time for
the density is required), a Luminosity $L_{bol}$, a photospheric velocity $v_{ph}$, and
a set of abundances. This is clearly a large parameter set. However, parameters
are not independent. $L_{bol}$ and $v_{ph}$ combine to determine the temperature $T(r)$.
In combination with the abundances, $T(r)$ and $Ab(r)$ determine the excitation
and ionization state of the gas. All these in turn determine line and electron opacities.
Relativistic zero-order corrections are included. Unlike other codes (\eg\ \citealt{Fisher00,Thomas11}), therefore,
we cannot arbitrarily change one value and leave all others unaffected. While this
makes the problem highly non-linear, it also makes the task of identifying the
range of likely optimal values easier. Here we discuss the procedure we adopted
in this particular case. 

We began with setting a density distribution.  We then established the best 
values of $L_{bol}$ and $v_{ph}$ at all epochs, and tested their uncertainty. We did
this keeping abundances constant as a function of velocity, to avoid introducing
too many free parameters. Changing the abundances in the ejecta may also alter the ionization and thermal conditions, due to changes in the opacity and hence the back scattering rate. 
However, using the same radiation transport code \citet{Mazzali08} examined the affects of varying the abundances in the ejecta. They found that the uncertainty on the abundances derived using this method is \ab20\%, and that the ``global'' parameters $L_{bol}$ and $v_{ph}$ are not significantly effected by changes in the abundances of this magnitude. 

\begin{figure*}
\centering
\includegraphics[scale=0.3]{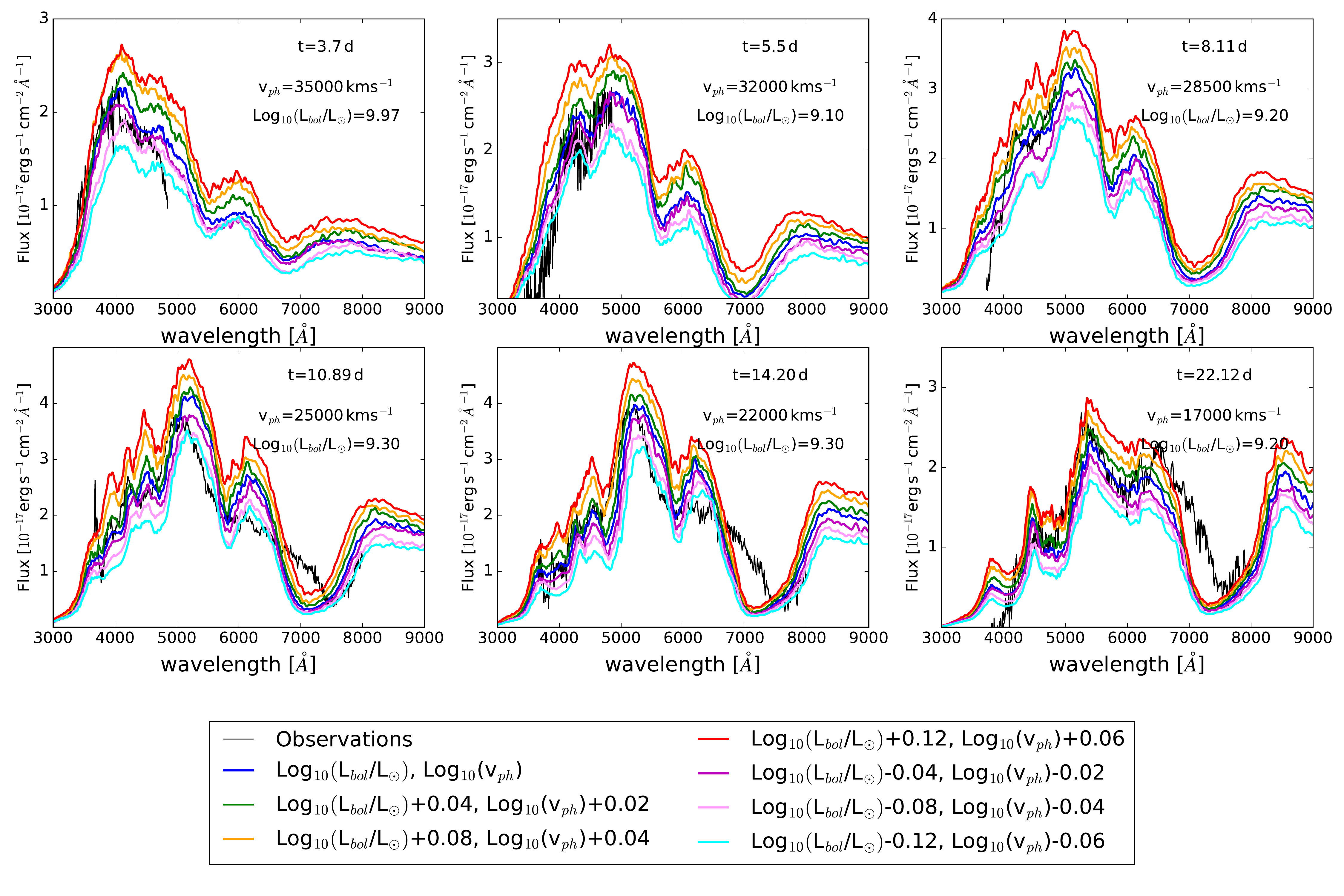}
\caption{Spectral observations and models of SN 2016jca with varying input 
luminosities and photosperic velocities, where $\frac {L_{bol}} {v_{ph}^2} \approx const$.
The blue model is the best fit model.}
\label{fig:LVdif}
\end{figure*}

\begin{figure*}
\centering
\includegraphics[scale=0.3]{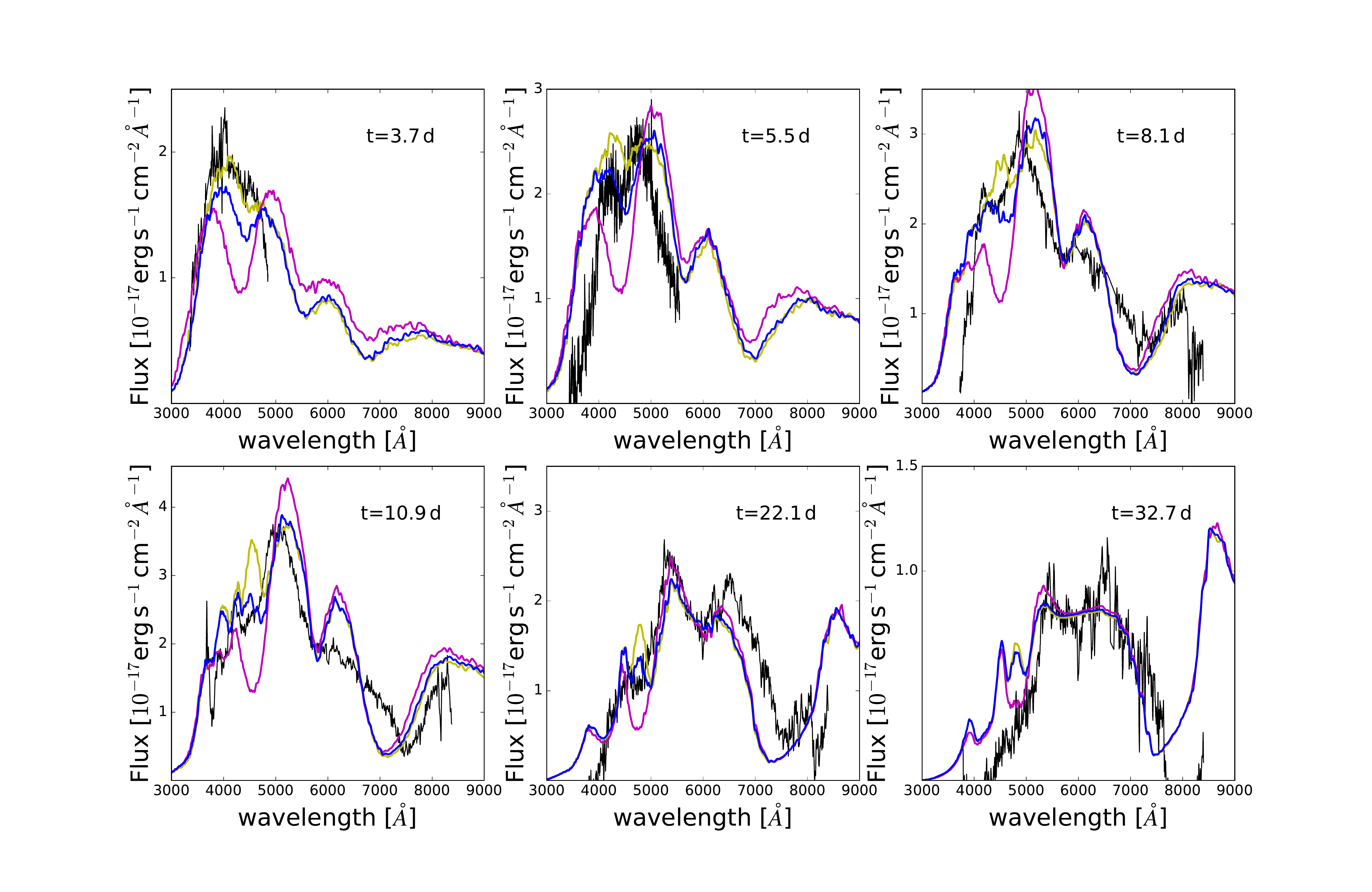}
\caption{Synthetic spectral models produced with no \Nifs, but different stable Fe mass fractions of
1$\times10^{-4}$ (blue), 1$\times10^{-5}$ (yellow), 1$\times10^{-3}$ (magneta).}
\label{fig:Fe}
\end{figure*}

\section{Explosion density profile}
SN\,2016jca has a fairly rapidly evolving light curve compared to other well-studied GRB-SN, which reach peak later. 
Therefore, to get a first estimate of the combination of 
\KE\ and \Mej, which produce the light curve and the spectra, we follow the work in \cite{Mazzali17} and use two well studied SNe, SN\,2006aj, and SN\,1998bw , as  starting points. SN\,2006aj (\Mej=2\Msun,  \KE=1$\times10^{52}erg$) \citep{Mazzali06} has a similar light curve suggesting the mass may be similar to that of SN\,2016jca.
 Whereas   SN\,1998bw (\Mej=10\Msun,  \KE=5$\times10^{52}erg$) \citep{Iwamoto98} has  similar spectra suggesting that the \KE/\Mej\ value is similar  to SN\,2016jca.
We consider both the time to peak and the width of the light curve. 
If we use SN\,2006aj as a starting point the two methods of deducing \Mej\ and \KE\ from the peak time and the light curve width yield different results, suggesting  SN\,2006aj  is not a good analogue. On the other hand, if we apply the same methodology on SN\,1998bw we get in both cases \Mej=6.5\Msun, and \KE=4$\times10^{52}erg$, suggesting SN\,1998bw has the correct  \KE/\Mej\ ratio, and SN\,2016jca had a smaller mass. Therefore we use  this  \KE/\Mej\ value as the basis for the spectral models below. 

The explosion model used for SN\,1998bw was CO138 \citep{Iwamoto98}.
It is the explosion of the $\sim$14\Msun\ CO core of a massive star with main 
sequence star mass \ab40\Msun \citep{Iwamoto98}.
In SN\,1998bw a high isotropic \KE\ ($\sim5\times10^{52}$erg) was used to match the 
observed broad absorption features and to produce the large amount of low velocity \Nifs\ 
typical of SN\,1998bw and of all other GRB-SNe.
A black hole remnant was assumed, and the inner 3\,\Msun\ were therefore excised,
resulting in an ejecta mass of \ab11\Msun. In this work we have used the unaltered hydrodynamic model from \citet{Iwamoto98}. Furthermore, when we scaled  the \Mej\ we kept the  same \KE/\Mej\ ratio.

Figure \ref{fig:5masses} shows models produced with 2\Msun, 4\Msun, 6.5\Msun, 8\Msun\ and 11\Msun. 
The same values of $L_{bol}$ and $v_{ph}$ were used to 
keep a constant $T(r)$. For all the models the abundances were scaled, at the expense of oxygen and neon with the same proportionally,  to keep roughly the same absolute mass of all other elements.
Therefore as the total mass of our ejecta decreases the mass of the heavier elements increases with respect to the light elements (i.e. oxygen and neon).
This is counter intuitive and may be uncertain as we get to masses further away from the 
original explosion model. However, the lowest masses are unlikely to produce  a synthetic light curve with the correct width.
It is apparent from the models that there is  not enough absorption in the
2 \Msun\ model and we can rule it out. Furthermore, the dilution factor, $w$, in the models should be \ab0.5. This is only the case for models near 6.5\Mej, and deviates significantly as the \Mej\ varies away from 6.5\Mej. Therefore the 6.5\Mej\ model produced the best results.  Combining this with information about the synthetic light curve model produced below, we take the range of acceptable values as models, and proceed with the 6.5$\pm$1.5\Msun. 

We note we did not test masses below 2\Msun, as such 
low masses would not be able to produce the width of the light curve of SN\,2016jca. In fact is is likely that the 4\Msun\ model would also not be able to reproduce the observed light curve.  As discussed below, we produce a model light curve to check our density profile and abundance structure and find good agreement with the our model and observed bolometric light curve.

\begin{figure*}
\centering
\includegraphics[scale=0.3]{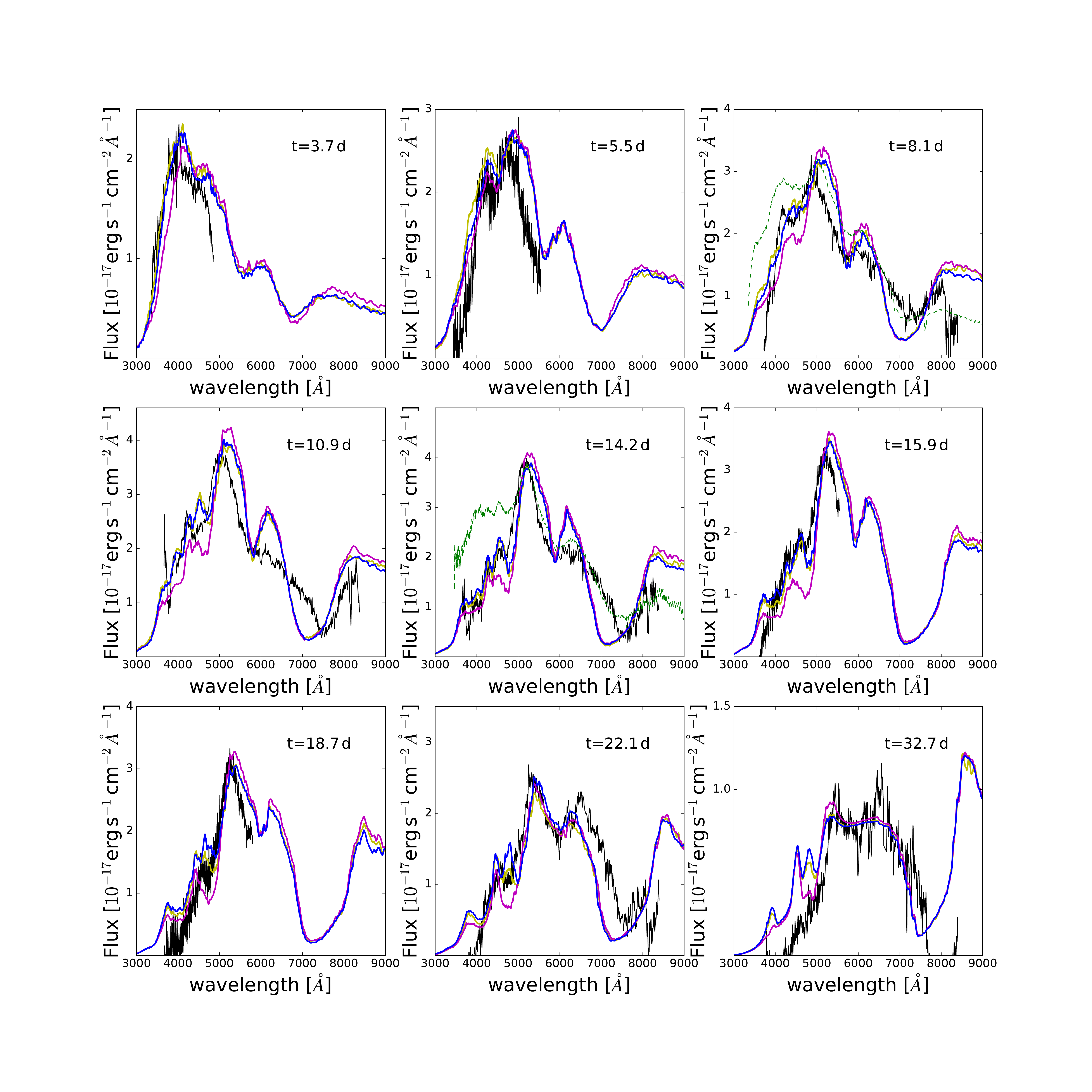}
\caption{Synthetic spectra produced with varying constant \Nifs\ abundances. 
The \Nifs\ abundances are \ab0.4\% (blue), \ab0.2\% (yellow), and \ab0.6\% (magneta).
At 8.1\,d and 14.2\,d the spectra of SN\,1998bw have been plotted (green dashed) normalized to the peak of the SN\,2016jca spectra. }
\label{fig:flatNi}
\end{figure*}

\section{Determining errors}
\subsection{Bolometric luminosity}

As we mentioned above, we used constant abundances throughout the ejecta to
determine $L_{bol}$ and $v_{ph}$ at the various epochs and their uncertainty from 
average models.  We use a set of models which have an elemental distribution that reflects the
composition of the collapse of a massive stellar carbon-oxygen core as a starting point. 
Using constant abundances as a function of velocity,  
for the \Mej=10\Msun\ model our most abundant elements 
are O(\ab70\%), Ne(\ab20\%) and C(\ab7\%), followed by Si(\ab1.5\%), S(\ab0.5\%) and \Nifs(\ab0.4\%), with the remaining 0.6\% consisting of Mg, Ca, Fe, Ti+Cr. 
It can be noticed that the \Nifs\ content far
exceeds the Fe abundance. We discuss this in the following sections.
This set of model offers a good 
reproduction of the spectral time series. 

To determine the errors, and to explain the fitting procedure, we computed a 
range of synthetic spectra starting from our main results but varying parameters 
($L_{bol}$, $v_{ph}$ and \Nifs\ abundance). We used values of $L_{bol}$ and $v_{ph}$ 
that yield correct flux levels and spectral distributions, and used abundances 
that yield reasonably accurate line strength for the most important features. 

The error on the bolometric luminosity,
was determined by producing models around the best fit model, with 
varying luminosities (see Figure \ref{fig:varyL}).
Changing $L_{bol}$ affects the overall flux, as can be see in the plot, but
also the temperature, such that models with lower $L_{bol}$ are redder, but this only
affects the flux region where most photon redistribution occurs, shortwards of
6000\,\AA. The combination of flux level and spectrum shape is what we used to
determine the best model and uncertainties. The plots show that on day 8 and
following epochs the error can be estimated as $\pm 5$\%, while at earlier times
it is larger, close to 10\% at day 5. At day 3 the flux is poorly determined, so we 
use a generous error of 25\%.

\subsection{Photospheric velocity}

Having determined $L_{bol}$ in our models, we now turn to $v_{ph}$. 
We produce models around a best fit model, with increasing and decreasing values 
of $v_{ph}$ (see Figure \ref{fig:difV}).

As changing $v_{ph}$ does not affect the total flux, the changes in the synthetic
spectra are due only to changes in temperature and opacity. The red part of the
spectra is largely unaffected, except for a shift of the edges of the broad
\OI/\CaII\ absorption caused by the different choices of photospheric velocity. 

The fits at 8.1 and 5.5\,d begin to become significantly worse when the
photospheric velocity is varied by 10\%, hence we adopt an error of 5\% in
$v_{ph}$ at 8.1 and 5.5\,d and later epochs. At 3.7\,d we adopt an error of 10\%
as the flux level is more uncertain here, and the shape of the hotter spectra (the
one with the lower  $v_{ph}$) matches the shape of the observations well.

\subsection{Bolometric luminosity vs photospheric velocity}

As we mentioned above, our input data are not independent. Looking at $L_{bol}$ and
$v_{ph}$, while individual error ranges have been assigned, only combinations of
parameters that roughly preserve temperature can keep an ionization and
temperature structure similar to the best fit model and yield reasonable spectra. 
Here we demonstrate that this is the case, combining results with +ve/-ve 
variations of $L_{bol}$ and $v_{ph}$ around the respective best values. Only values 
where $\frac {L_{bol}} {v_{ph}^2} \approx const$ reproduce the spectral shape. The set of models thus
produced are all similar in shape, except for the differences in line width and
strength resulting form using different depths, but the change in $L_{bol}$ causes the
overall flux  level to deviate, making it easy to discriminate among models and
leaving the actual flux calibration of the data as the main uncertainty.  
This is very useful in constraining the possible range of parameters, see Figure
\ref{fig:LVdif}. Therefore, if changing $L_{bol}$ to a larger value, only larger values
of $v_{ph}$ are compatible, and vice-versa, which reduces the allowed space of
parameters.

It is apparent from the models that there is no degeneracy between $L_{bol}$ and
$v_{ph}$, such that varying $L_{bol}$ by 10\% and $v_{ph}$ by 5\%, even in the
same direction, produces significantly worse fits after 5.5d. At
3.7\,d several models yield good fits, hence our errors here are of the order of
20\% on $L_{bol}$ and 10\% on $v_{ph}$. 

\begin{figure*}
\centering
\includegraphics[scale=0.3]{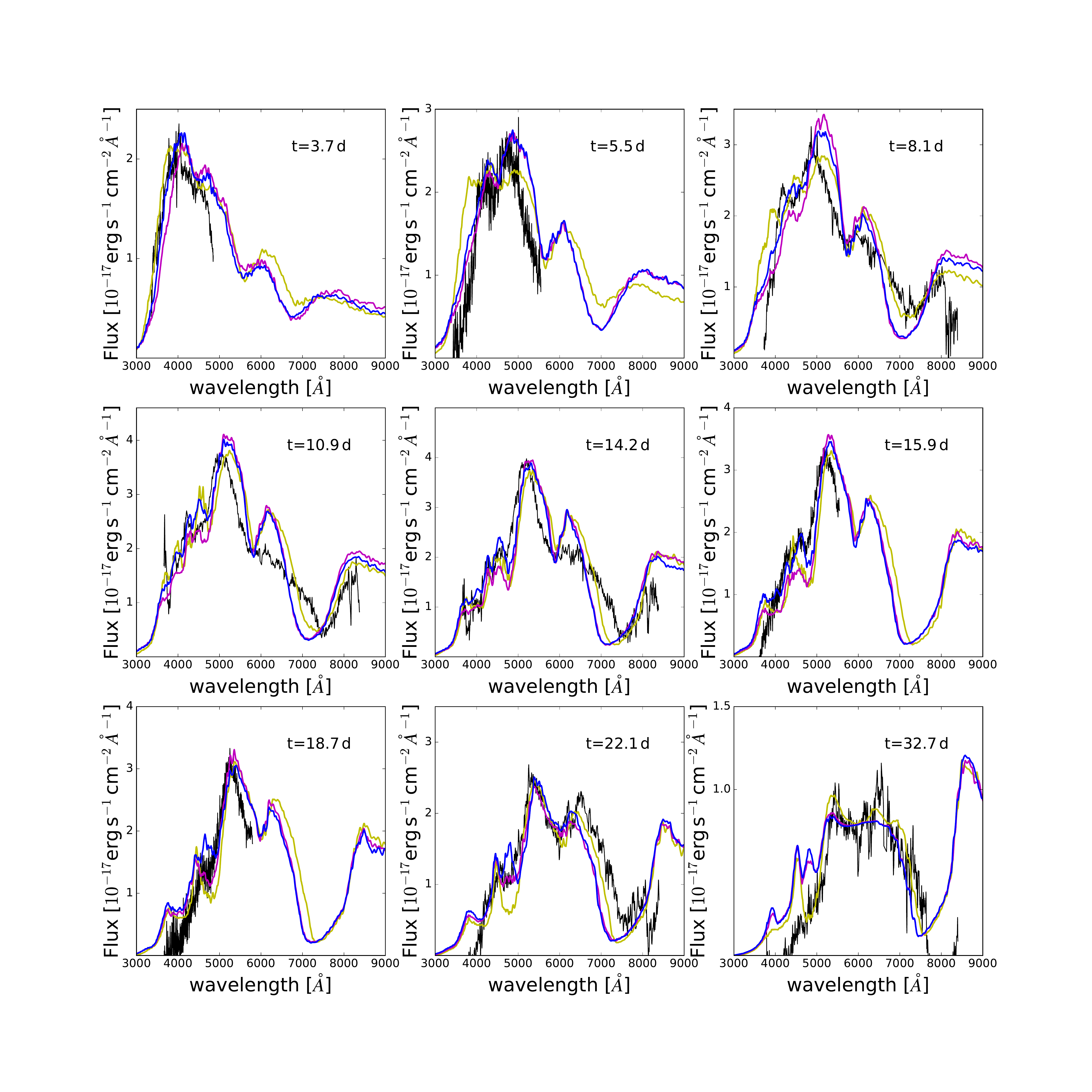}
\caption{Spectral models produced with \Nifs\ abundances. 
The best fit blue model has a \Nifs\ abundance of \ab0.4\%, the magenta model has a \Nifs\ abundance which decreases as velocity decreases, and the yellow model has the opposite trend. Note that a variation of this plot is presented in \citet{Ashall19}.}
\label{fig:stratNi}
\end{figure*}

\begin{figure}
\centering
\includegraphics[scale=0.6]{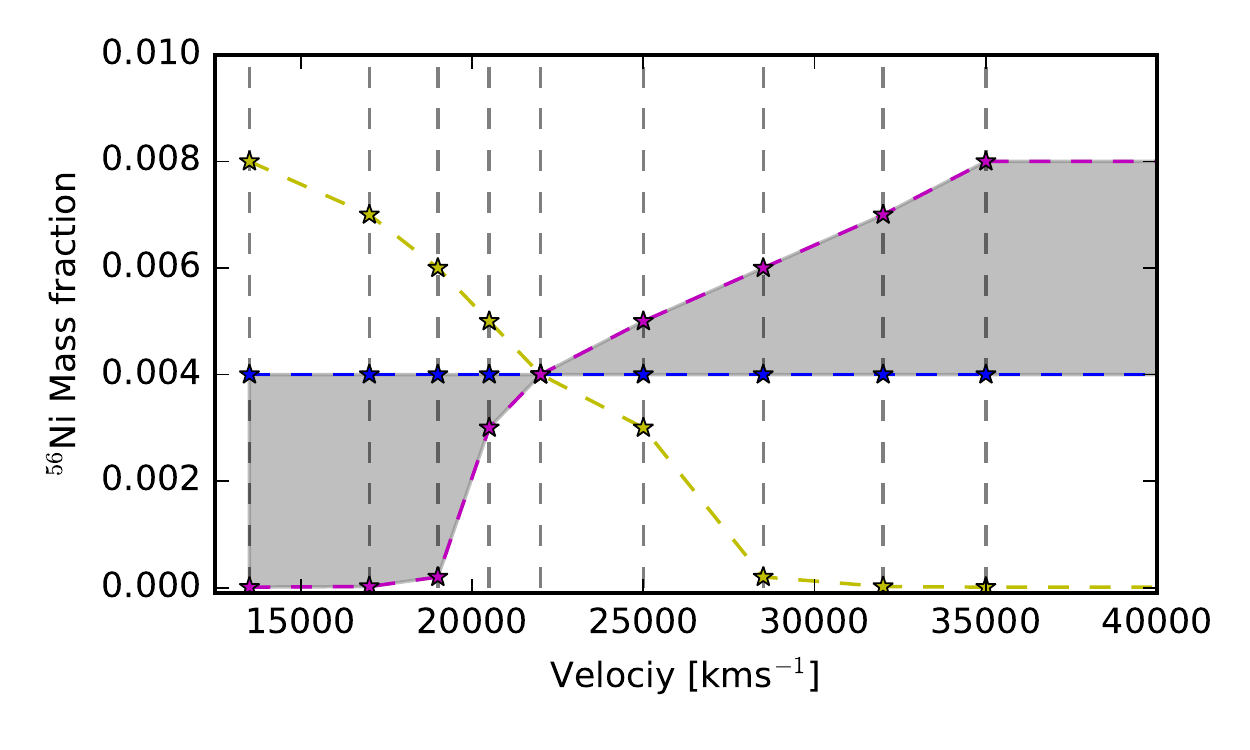}
\caption{ The \Nifs\ abundance distribution as a function of velocity for the three models presented in Figure \ref{fig:stratNi}. The values of \Nifs\ which produce good fits are highlighted in grey. The dashed horizontal lines represent the photospheric velocity from the nine spectral models. Note that a variation of this plot is presented in \citet{Ashall19}.}
\label{fig:stratNiab}
\end{figure}

\begin{table}
 \centering
 \caption{Input parameters for the spectral models, the errors are give in parenthesis. }
  \begin{tabular}{ccc}
  \hline
Epoch &   $v_{\rm ph}$   &  $L_{bol}$   \\
(rest-frame days)& km\,s$^{-1}$& log($L_{bol}$/L$_{\odot}$) \\
  \hline
  3.73& 35000(3500)&8.97(0.1) \\
  5.52& 32000(1600)&9.10(0.05) \\
 8.11& 28500(1425)&9.20(0.02) \\
10.89& 25000(1250)&9.30(0.02) \\
14.20& 22000(1100)&9.30(0.02)\\
15.96& 20500(1025)&9.26(0.02) \\
18.71& 19000(950)&9.26(0.02) \\
22.12& 17000(850)&9.20(0.02) \\
32.65& 13500(675)&9.00(0.02) \\
  \hline
\end{tabular}
\label{table:logofmodel}
\end{table}

\section{Determining the Ni distribution in the outer layers.}

 In Type I SNe most of the opacity is usually caused by metal lines, in particular
 Fe lines. GRB-SNe in general, and SN\,2016jca is no exception, are very red at
 early phases, although they are more luminous on average than other SNe\,Ic \citep{Prentice16}. 
 This red colour is often attributed to the high velocities reached in these SNe, but
 the effect is not simply one of temperature. It is again the result of different
 line opacities.  Here we show that the early red colour is due to the presence of
 \Nifs\ at high velocities. This makes GRB-SNe different from low-energy SNe\,Ic,
 which tend to have blue spectra at early times \eg SN\,1994I \citep{Sauer06}.
 First we test the iron abundance. 
 Although it should be expected that the Fe content at early times should reflect 
 that of the progenitor star, and therefore presumably be similar to that of the 
 host galaxy, we ask the question whether the red colour is due to Fe abundance by testing models with varying Fe composition.

Figure \ref{fig:Fe} shows synthetic spectra for the time series of spectra of
SN\,2016jca obtaining for different Fe abundances. These models have been produced
using stable Fe abundances, and with no \Nifs, so the Fe content does not change as a function
of time. For all spectra at 5.5, 8.1, 10.9 day there is excess flux in the blue part of the
spectra. Furthermore, for large Fe abundances strong \FeII\  lines ($\lambda \lambda$ 4923 5018 5169), which are much deeper than the observations appear in the 4000-5000\AA\ range. Even for large Fe abundances the blue part of the spectrum is not
sufficiently blocked, as there are not enough strong Fe lines in this region.  
Co and Ni
are required to block the excess flux in the blue part of the spectra, and as no
stable Co is produced in GRB-SN explosions, high velocity \Nifs\ is needed in the
models.

As Fe cannot shape the spectra of SN\,2016jca, we show here that \Nifs\ and its
decay products offer the correct opacities. We tested the abundances of various
elements to verify what causes the spectra to be red, and found that the element
that most affects the shape of the spectrum is \Nifs. We show a set of
models with different \Nifs\ abundances and a flat distribution. These clearly
show that \Nifs\ and decay products have a major effect on the spectral shape, especially at early times.
More \Nifs\ suppresses the blue and enhances the redder part of the spectrum by
fluorescence, in particular near the $V$ band, where line opacity is smaller and
photons can escape. We show models computed with different constant initial \Nifs\
abundances in Figure \ref{fig:flatNi}. It is apparent that the best fit \Nifs\ 
abundance is 0.4\%, the 0.2\% model does not have enough UV blocking at day 5.5 or 8.1  and the 0.6\% model has too much Fe absorption (produced from the decay of \Nifs) at later times (e.g. 15.9 and 22.1\,d).  
In fact, the results from Figure \ref{fig:flatNi} may suggest that 
at early times (high velocities) there is a large \Nifs\ abundance which can
block the flux in the UV,  
and at later times (lower velocities) the \Nifs\ abundance is much less.

Having a constant mass fraction of \Nifs\ as a function of velocity, is difficult to coincide with an 
aspherical explosion, especially if the \Nifs\ is produced on the side of the GRB-jet. 
However, it could be expected that the \Nifs\ abundances could decrease as a function of 
decreasing velocity. This is as the photosphere recedes the jet, heavily synthesized region surrounding the 
jet, will becomes a smaller overall fraction of the total observed region. Although in the center of the ejecta 
the \Nifs\ abundances would be expected to increase again. 
To test for this models were produced for both increasing and decreasing \Nifs\ abundance as a function of velocity, see Figure \ref{fig:stratNi}.

The yellow model in which the \Nifs\ abundance increases as a function of decreasing velocity produces poor fits at both early and late times. 
At later times (22.12\,d) there is too much absorption at $\sim$4200\AA, caused by the decay of \Nifs, and at early times (3.7, 5.5 and 8.1\,d). 
This is as there is not enough blocking by \NiII\ and \CoII\ lines. 
The blue models have a constant \Nifs\ abundance of  0.4\% and produce good fits at all epochs. 
However, the magenta models which have a decreasing \Nifs\ abundance as a function of decreasing velocity
 produce fits which are better or as good as the blue models.
Therefore, the range of best fit lines between the  magenta and blue models.
This range of \Nifs\ abundance is shown in the Figure \ref{fig:stratNiab} as the shaded region. 

The fact that we have a solution where the \Nifs\ abundance is decreasing as velocity could imply that 
 SN is aspherical both in shape and in elemental distribution. This is as the \Nifs\ could have been sythesised
on the side of the jet. As time passes the jet, and heavily synthesized material surrounding it, 
would cover a smaller overall fraction of the total material observed above the photosphere, and more 
lighter material on the side of the ejecta would be observed. Hence the metal abundance would decrease
at the expense of lighter elements. 

\citet{Izzo19} have also claimed evidence for high-velocity \Nifs\ in a GRB-SN, they suggest that this could have been produced by a jets and a cocoon breaking through a stellar surface.  However,  spectral models of the radial dependence of \Nifs\ was first suggested by the original preprint of \citet{Ashall19}. 
Both pieces of work use radiation transport codes, however \citet{Izzo19} use a  code based on the one from this work and 
did not perform an error analysis. They claim that that SN\,2017iuk requires a flat 
density structure in the outer most layers. 
In this work and in \citet{Ashall18a}
it was found that there was no need to enhance the outer density structure in the outer layers for SN\,2016jca, and an increase in abundance of Fe-group elements was sufficient. 
This was not tested by \citet{Izzo19}, and it could in fact be an alternative solution.
In the case presented here   blobs of \Nifs\ could have been  dredged up from the centre of the explosion by the jet. However, the need for high velocity \Nifs\ may not be required in all SNe, as is seen in Figure \ref{fig:flatNi}, SN\,1998bw has  more flux in the blue than SN\,2016jca. This is likely to be caused by a lack of Fe-group elements at high velocity in the ejecta of SN\,1998bw, which could have been due to the even being observed slightly off axis, hence SN\,1998bw had a weak afterglow.

\section{Bolometric light curve}
To verify that the results we obtained are realistic we produce a model bolometric light curve of SN\,2016jca. The light curve code is a Monte Carlo code which was first presented in \citet{Cappellaro97} and then expanded in \citet{Mazzali2000}.
In the light curve models gamma-ray opacity is assumed to be constant and equal to 0.027 $cm^2 g^{-1}$  \citep{Sutherland84}. Positrons are propagated using  an opacity of 7 $cm^2 g^{-1}$ as an approximation \citep{Sutherland84}. Optical photons are supposed to encounter an opacity dominated by line opacity and which therefore depends on composition and slowly decreases with time, as in (\citealt{Hoeflich96}; \citealt{Mazzali01}.) 
 
\begin{figure}
\centering
\includegraphics[width=\columnwidth]{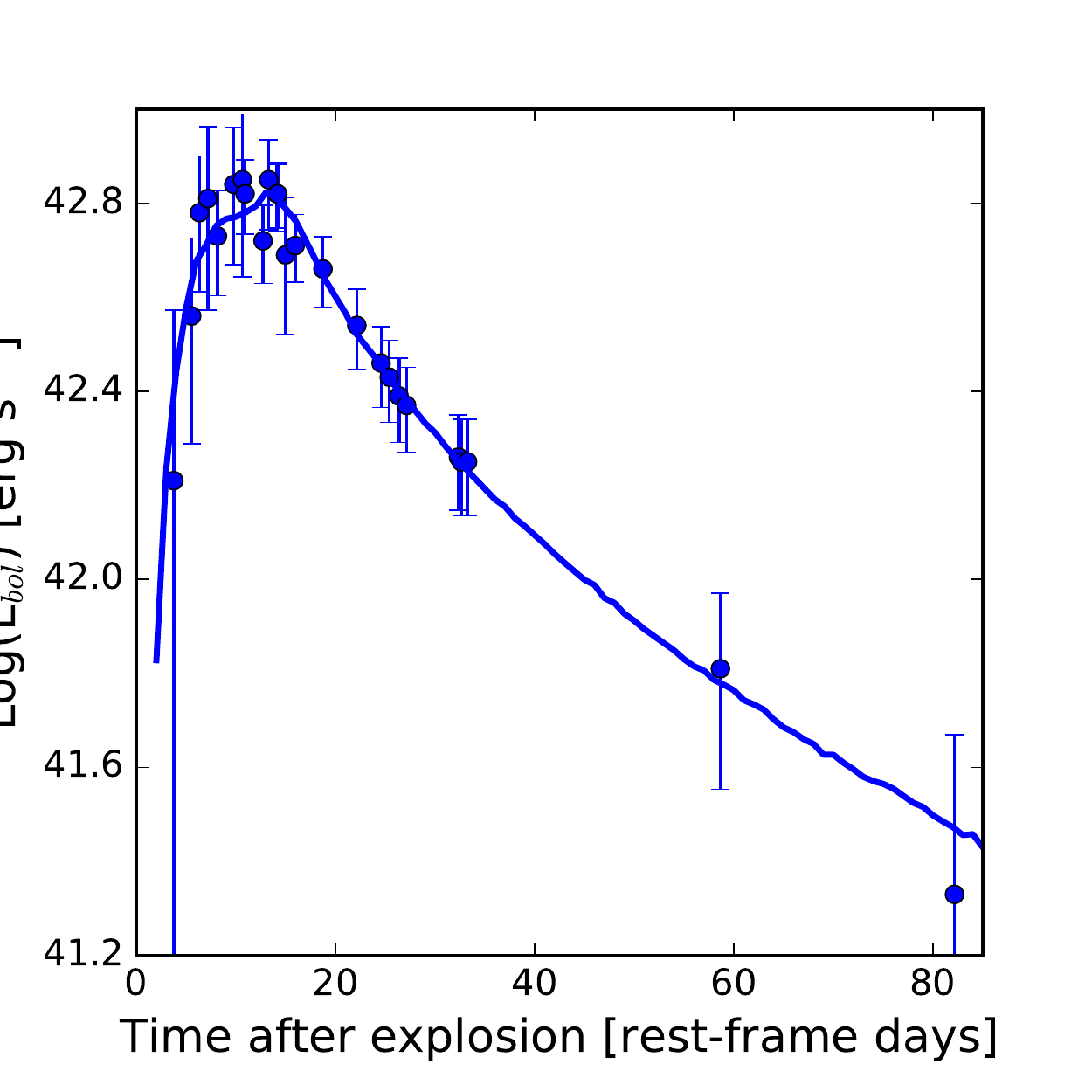}
\caption{ The observed (blue circles) and modeled (blue solid line) bolometric light curves of SN\,2016jca.}
\label{fig:BOL}
\end{figure}

 With a given density structure and abundance composition the code follows the propagation of photons through the SN atmosphere in the expanding ejecta and a synthetic light curve is produced. The abundance are assumed to be constant below 13500\kms\ and the density profile from the 
hydrodynamical model is used. The bulk of \Nifs\ is assumed to be centrally located, and the total mass of the \Nifs\ was taken as 0.27\Msun\, as derived from the peak of the bolometric light curve \citep{Ashall19}.  
Figure \ref{fig:BOL} shows the observed bolometric light curve from \citet{Ashall19} and the synthetic light curve. The model provides a consistent fit to the observations, which demonstrates that the input density profile, ejecta mass, \KE\ and \Nifs\ distribution is consistent with both the spectra and light curves of SN\,2016jca.  

\section{Conclusions}
Analysing the spectra of SN associated with LGRBs is one of the most promising ways to 
obtain physical information about these rare explosions. 
However, previous studies have tended to just publish the best fitting results with little 
information about the fitting method, or associated errors.
Therefore in this analysis, we concentrate on the method, errors, and results. 

As an example, a high quality time series of the photospheric phase 
spectra of SN\,2016jca was analysed using a Monte Carlo radiation transport code.
It was demonstrated how changes in $L_{bol}$ affects the temperature in the models, as well as the 
overall flux, where models with lower $L_{bol}$ are redder. This occurs 
in the flux region where most photon redistribution occurs, short wards of
6000\,\AA. Whereas, changing $v_{ph}$ affects the temperature and opacity.
For most epochs (where the GRB afterglow is weak) the error in $L_{bol}$ and $v_{ph}$ is 5\%. 

It was also shown when $\frac {L_{bol}} {v_{ph}^2} \approx const$ spectral fits begin to degrade if
parameters are moved away from the best fit solutions.
Even varying the $L_{bol}$ by 10\% and $v_{ph}$ by 5\%, so the parameters
roughly preserve temperature and can keep a similar ionization and
temperature structure, produces worst fits. This demonstrates
for a unique set of observations there is only one $L_{bol}$ and $v_{ph}$ which
produce good fits. 

The affect of vary ejecta mass was also explored. Previous work has examined the
slope of the outer density profile and and how it affects the broadness of the features
\citep{Mazzali17}.
Here we concentrated on having a constant \KE/\Mej\
but varying the over all \Mej. This roughly keeps the spectral shape constant, 
as long as their is enough 
mass above the photosphere to form opacity and the absolute mass of the elements which form the lines 
stays constant between different values of \Mej. 
By moddeling only the photospheric phase spectra we found that the errors on 
\Mej\ were 20\%, and because of this so were the errors on \KE. However for an individual 
\Mej\ the error on \KE\ is much smaller (see \citet{Mazzali17}). 
Furthermore, simultaneous photopsheric phase, nebular phase, and bolometric light curves models 
would dramatically reduce the error on \Mej. 
In the case of SN\,2016jca the best fit density profile was 
6.5$\pm$1.5\Msun, and a \KE\ of 4$\pm$0.8 $\times 10^{52}$\,erg (1-3$\times 10^{52}$\,erg when 
correcting for asphericity; \citealt{Maeda02}).

In the modeling procedure the initial abundances were kept constant and 
consistent with explosion models. 
However, as each individual SE-SN is unique it is likely that the abundances are not always the same, which can add a lot more free parameters. Therefore, the method which is
used is to varying individual abundances by a constant value for all epochs to see if the fits improve. 
For our test case, SN\,2016jca, the observations demonstrated that there was less flux in the blue part
of the spectra compared to previous events, possibly due to increased line blanketing. It was found that increasing the 
stable Fe abundance did not produce the correct opacity. A time variable opacity was required, and 
high velocity \Nifs\ provided this. 
However having a constant \Nifs\ abundance at high velocities is hard to coincide with explosion models, 
therefore a decreasing \Nifs\ abundance as a function of decreasing velocity was also tested, and 
these models produced marginally better fits to the data. In this scenario the \Nifs\ would have been
sythesised around the jet as it passed through the stellar surface, or it could have been dredged up from the centre of the star by the jet. Either way as time passes and the photosphere recedes the jet and the high energy material surrounding it would contribute to less and less over the overall fraction of the observed material. Hence, the \Nifs\ abundance would decrease as a function of  decreasing velocity. 

The work presented here has demonstrated how radiation transport models can be used to obtain
physics information about SE-SN explosions. We have specifically concentrated on the fitting procedure 
and the associated errors with them. Although the problem is non-linear it is apparent that only 
one set of parameters produce the best fitting synthetic models. 
This paper can be  used as a guideline for future studies 
which use the same or similar radiation transfer codes.

\section*{Acknowledgements} 
Chris Ashall acknowledges the support provided by the National Science Foundation under Grant No. AST-1613472.
Work in this paper was based on observations made with ESO 
Telescopes at the Paranal Observatory under programmes ID  098.D-0055(A), 098.D-0218(A) and 298.D-5022(A).






\bibliographystyle{mnras}
\bibliography{BB}




\bsp	
\label{lastpage}
\end{document}